\documentclass[pdftex,useamsfonts]{pasj01}
\usepackage{graphicx}
\usepackage{color}
\usepackage{natbib}

\let\leftroot\relax
\let\uproot\relax

\usepackage[fleqn]{mathtools}
\begin{document}

\title{Impact of the initial disk mass function on the disk fraction}
\author{Ryou Ohsawa$^1$, Takashi Onaka$^2$, Chikako Yasui$^2$}
\email{ohsawa@astron.s.u-tokyo.ac.jp}
\affil{$^1$Institute of Astronomy, University of Tokyo, 2-21-1 Osawa, Mitaka, Tokyo 181-0015, Japan}
\affil{$^2$Department of Astronomy, Graduate School of Science, The University of Tokyo, 7-3-1 Hongo, Bunkyo-ku, Tokyo 113-0033, Japan}
\maketitle

\begin{abstract}
  The disk fraction, the percentage of stars with disks in a young cluster, is widely used to investigate the lifetime of the protoplanetary disk, which can impose an important constraint on the planet formation mechanism. The relationship between the decay timescale of the disk fraction and the mass dissipation timescale of an individual disk, however, remains unclear. Here we investigate the effect of the disk mass function (DMF) on the evolution of the disk fraction. We show that the time variation in the disk fraction depends on the spread of the DMF and the detection threshold of the disk. In general, the disk fraction decreases more slowly than the disk mass if a typical initial DMF and a detection threshold are assumed. We find that, if the disk mass decreases exponentially, {the mass dissipation timescale of the disk} can be as short as $1\,{\rm Myr}$ even when the disk fraction decreases with the time constant of ${\sim}2.5\,{\rm Myr}$. The decay timescale of the disk fraction can be an useful parameter to investigate the disk lifetime, but the difference between the mass dissipation of an individual disk and the decrease in the disk fraction should be properly appreciated to estimate the timescale of the disk mass dissipation.
\end{abstract}

\section{Introduction}
\label{sec:intro}
Protostars are surrounded by disks consisting of gas and dust (referred to as a protoplanetary disk), which have been investigated as sites of planet formation. Young stellar objects (YSOs) of low- and intermediate-mass, systems consisting of a protostar and a protoplanetary disk, are roughly categorized into three classes by their spectral energy distributions (SEDs). Class I objects are the youngest with infrared emission from dust dominating the SED; Class II objects have intermediate ages with the SEDs characterized by a combination of stellar and hot dust emission; Class III objects are the oldest, consequently showing the weakest dust emission \citep[e.g.,][]{williams_protoplanetary_2011,bachiller_bipolar_1996}. The mass and surface density of protoplanetary disks decrease with time \citep{wyatt_origin_2007,wyatt_evolution_2008}, whose mechanism is still under debate. Although the dissipation involves a number of processes, photoevaporation likely plays an important role, especially for low-mass stars \citep{armitage_dynamics_2011}. Planets are thought to form in the protoplanetary disk. The dust and gas surface densities of the protoplanetary disk are important parameters that determine the accretion rates to the cores and the final mass of gas giant planets to form \citep[e.g.,][]{ida_toward_2004}. It is therefore important to understand the evolution of the protoplanetary disk for the investigation of planet formation.

Different regions within a protoplanetary disk can be investigated with different wavelengths. Continuum emission in the near-infrared comes mostly from hot dust in the innermost part of the disk. Emission in the mid-infrared comes either from the surface of the disk or the mid-plane at the disk at a distance of several AUs. The far-infrared emission is dominated either by the mid-plane or the outermost part of the disk \citep{dullemond_inner_2010}. Atomic and molecular emission lines trace gas components and accretion activities. The investigation of the planet-forming disk evolution requires observations in the near- or mid-infrared.

There are, however, difficulties in the observational investigation of the evolution of protoplanetary disks. First, the age of individual YSOs is difficult to estimate. It can be estimated from the loci in the Hertzsprung-Russel diagram \citep[e.g.,][]{strom_circumstellar_1989,kenyon_pre-main-sequence_1995} or by near-infrared and X-ray photometry \citep{getman_age_2014}. However, these have inherent non-negligible uncertainties. The estimate based on the equivalent width of absorption lines has a better accuracy \citep{takagi_age-determination_2010,takagi_age_2011}, but it requires high-resolution spectroscopy, which is not always available. The mass of protoplanetary disks is also difficult to estimate from near- and mid-infrared observations since they are optically thick at these wavelengths.

The disk fraction, defined as the fraction of stars with disks in a star cluster, is thus widely used to study the evolution of protoplanetary disks since it relies on the age of star clusters, which reduces the uncertainties in the ages of individual stars, and since the presence of a disk can be estimated directly from the loci in the color-color diagram of near- and mid-infrared photometry as excess emission, although it does not estimate the disk mass. \citet{haisch_disk_2001} show that the disk fraction estimated in the near-infrared decreases gradually with the cluster age, while \citet{mamajek_initial_2009} reports that the disk fraction decreases exponentially with the time constant of $2.5\,$Myr (decay timescale of the disk fraction, hereafter DTDF).

The DTDF may change with the stellar mass and metallicity of the cluster \citep[e.g.,][]{yasui_lifetime_2009}. \citet{hernandez_herbig_2005} derive the disk fraction of nearby OB associations using the $JHK$-bands and suggest that the disk fraction for intermediate-mass stars is lower than that for low-mass (${\lesssim}1\,M_\odot$) stars. The result indicates the fast dissipation of the inner-disk for intermediate-mass stars. A similar result is reported by \citet{carpenter_evidence_2006}, based on the \textit{Spitzer} observations of U Sco OB association. \citet{kennedy_stellar_2009} investigate the disk fraction of nine clusters, using infrared excess estimated from \textit{Spitzer} observations and H$\alpha$ equivalent width, and show that the disk fraction decreases with increasing stellar mass. \citet{hernandez_spitzer_2008}, using 2MASS and \textit{Spitzer} data, find that the disk fraction of $\gamma$ Vel cluster is smaller than other clusters at a similar age ($\simeq 5\,$Myr). They suggest that the small disk fraction may be attributed to the strong radiation field from massive stars in the cluster. For low-mass ($\lesssim 2\,M_\odot$) stars, the disk fractions estimated at the near- and mid-infrared do not show any appreciable difference between each other. On the other hand, \citet{yasui_rapid_2014} suggest that the disk disappears by $\sim 3$Myr faster at the near-infrared than at the mid-infrared for intermediate-mass ($\sim $1.5--7$\,M_\odot$) stars, being consistent with the lack of planets discovered in the vicinity of the intermediate-mass stars. \citet{ribas_disk_2014} investigate the disk fraction of 22 young clusters using SED fitting from optical to mid-infrared. They show that the primordial disk detected at shorter (3.4--12$\,\mu$m) wavelengths disappears faster than that at longer (22--24$\,\mu$m) wavelengths, indicating that the disk closer to the star is evolving more rapidly.

Although the evolution of the disk can be studied efficiently by the disk fraction, the DTDF is not equal to the mass dissipation timescale of an individual disk (MDTID). The DTDF is a statistical measure and does not directly indicate the variation in the disk mass of individual objects. It is the timescale of the disk dissipation that governs the planet formation process. However, the reliability of the DTDF as an estimator of the MDTID has not been studied systematically. It is thus of importance to understand the relation between the DTDF and the MDTID in a semi-quantitative manner.

How the disk fraction decreases with the cluster age depends on the statistical properties of the cluster. \citet{owen_protoplanetary_2011} investigate the photo-evaporation mechanism of the protoplanetary disk by X-rays with radiation-hydrodynamic simulations. They demonstrate that the observed disk fraction at the near-infrared can be reproduced by the X-ray luminosity function of the cluster. Therefore, the behavior of the disk fraction must be investigated taking account of the distribution of physical parameters of the cluster. Sub-millimeter observations indicate that the total dust mass in the disk is distributed over a wide range \citep[e.g.,][]{andrews_circumstellar_2005,andrews_submillimeter_2007}. \citet{armitage_dispersion_2003} show that the observed spread of the disk lifetime in the Taurus cloud is consistent with theoretical models of disk evolution if a dispersion in the initial disk mass is assumed. The initial disk mass function (DMF) can thus affect the time variation of the disk fraction in a statistical way, but this effect has not been investigated in detail yet.

In this paper, we devise a simple model to estimate the effect of the initial DMF on the time variation of the disk fraction. The difference between DTDF and MDTID is calculated, taking account of the broadness of the DMF. Details of the model are described in Section~\ref{sec:model}. In Section~\ref{sec:discussion}, the variation in the calculated disk fraction is investigated and compared with the observed disk fraction. The results are summarized in Section~\ref{sec:conclusion}.

\section{Model}
\label{sec:model}
The evolution of the disk fraction is investigated by taking into account the disk mass distribution. In the present model, we consider the observation of a young star cluster in the near- or mid-infrared, where the protoplanetary disk is optically thick. In observations, the target stars are selected by their spectral types and thus the stellar mass is in a narrow range. In the following, we do not consider the distribution of the mass of the central star. Infrared excess ($f^{\rm ex}$) is assigned to each star. The star is recognized as being associated with a protoplanetary disk when $f^{\rm ex}$ is larger than a critical value ($f^{\rm th}$). All stars in the cluster are assumed to be born as a single star at the same time with different disk masses. The effect of the age dispersion within the cluster is discussed in Section~\ref{sec:dis:justification}. The disk fraction is defined by the fraction of such stars with $f^{\rm ex} > f^{\rm th}$ in the cluster.

The DMF, the fractional number of the disk with the mass between $m_{\rm d}$ and $m_{\rm d}{+}{\rm d}m_{\rm d}$, is defined by $\phi_0(m_{\rm d}){\rm d}m$ and $\phi_0(m_{\rm d})$ is normalized as
\begin{equation}
  \int \phi_0(m_{\rm d}){\rm d}m_{\rm d} = 1.
  \label{eq:mass_distribution}
\end{equation}
Based on a survival analysis, \citet{andrews_circumstellar_2005} show that the DMF of the Taurus-Auriga region is well approximated by a log-normal distribution. Similar results are also reported by \citet{andrews_submillimeter_2007} and \citet{mann_submillimeter_2010}. According to their results, the initial DMF is approximated by
\begin{align}
  \label{eq:massfunction_logn}
  \phi_0&(m_{\rm d}){\rm d}m_{\rm d} = \nonumber\\
  &
  \frac{1}{m_{\rm d}\sqrt{2\pi}\sigma}
  \exp\left[
    -\frac{1}{2}
    \left(\frac{\log{m_{\rm d}} {-} \log{\mu}}{\sigma}\right)^2
  \right]{\rm d}m_{\rm d},
\end{align}
where $\mu$ and $\sigma$ define the location and width of the distribution.

The disk dissipates with time. \citet{luhman_disk_2010} suggest that the inner optically thick disk evolves into an optically-thin phase rapidly because the number of transitional disks may be small. The evolution of the inner disk mass is, however, not well understood. If the disk is regarded as an isolated system, the mass loss rate of the disk should be determined by the parameters of the disk itself, such as the disk mass, the luminosity of the central star, and the angular momentum. To simplify the case, we assume that the dissipation of the disk mass is given by an unary function of the disk mass:
\begin{equation}
  \label{eq:disk_dissipation}
  \frac{{\rm d}m_{\rm d}}{{\rm d}t}
  = - \frac{1}{\zeta(m_{\rm d})},
\end{equation}
where $\zeta(m)$ is an arbitrary positive definite function to define the disk mass loss rate. By integrating Equation~(\ref{eq:disk_dissipation}), the relationship between $m_{\rm d}(t)$ and $t$ is obtained.
\begin{equation}
  \label{eq:dmdt_integrated}
  \int^{m_{\rm d}(t)}_{m_{\rm d}(0)} \zeta(m) {\rm d}m = -t.
\end{equation}
We define $\mathcal{Z}(m)$ as the integration of $\zeta(m)$.
\begin{equation}
  \label{eq:Z_defined}
  \mathcal{Z}\left(m_{\rm d}(t)\right) - \mathcal{Z}\left(m_{\rm d}(0)\right)
  = -t.
\end{equation}
Since $\zeta(m)$ is positive definite, the integrated function $\mathcal{Z}(m)$ is a monotonically increasing function, for which the inverse function $\mathcal{Z}^{-1}(m)$ is uniquely defined. Thus $m_{\rm d}(t)$ is solved as
\begin{equation}
  \label{eq:mt_solved}
  m_{\rm d}(t)
  = \mathcal{M}\left(t,m_{\rm d}(0)\right)
  = \mathcal{Z}^{-1}\left(
    -t + \mathcal{Z}\left( m_{\rm d}(0) \right)
  \right).
\end{equation}
The inverse function $\mathcal{Z}^{-1}(m)$ is also a monotonically increasing function. $\mathcal{M}\left(t,m_{\rm d}(0)\right)$ monotonically decreases with $t$:
\begin{equation}
  \label{eq:mt_with_time}
  t > t' \iff
  {\mathcal M}\left(t, m_{\rm d}(0)\right) < {\mathcal M}\left(t', m_{\rm d}(0)\right).
\end{equation}
The MDTID is defined as a typical timescale of the decrease in ${\mathcal M}\left(t, m_{\rm d}(0)\right)$ as discussed in Section~\ref{sec:dis:behavior}. Instead, $\mathcal{M}\left(t,m_{\rm d}(0)\right)$ is an increasing function in terms of $m_{\rm d}(0)$:
\begin{equation}
  \label{eq:mt_with_mass}
  m_{\rm d}(0) > m'_{\rm d}(0) \iff
  {\mathcal M}\left(t, m_{\rm d}(0)\right) > {\mathcal M}\left(t, m'_{\rm d}(0)\right).
\end{equation}
Equation~(\ref{eq:mt_with_mass}) assures that the mass dissipation process does not change the order of the disk mass. Equation~(\ref{eq:Z_defined}) can be solved in terms of $m_{\rm d}(0)$:
\begin{equation}
  \label{eq:m0_solved}
  m_{\rm d}(0)
  = \mathcal{Z}^{-1}\left(
    t + \mathcal{Z}\left( m_{\rm d}(t) \right)
  \right)
  = \mathcal{M}\left(-t,m_{\rm d}(t)\right).
\end{equation}
The relationship between the initial DMF and the DMF at $t$ is given by
\begin{equation}
  \label{eq:diskmassfunction_with_age}
  \phi_t\left( m_{\rm d} \right) {\rm d}m_{\rm d}
  =
  \phi_0\left(
    \mathcal{M}\left(-t,m_{\rm d}\right)
  \right) {\rm d}\mathcal{M}\left(-t,m_{\rm d}\right).
\end{equation}
The DMF changes in time depending on the time evolution of $\mathcal{M}(-t,m_{\rm d})$. Although we assume that the initial DMF is given by a log-normal distribution, the shape of the DMF at $t$ can be different from a log-normal function. Here we discuss a general case of $\mathcal{M}(-t,m_{\rm d})$.

The near-infrared excess is directly related to the inner disk mass, not the total disk mass. The mass function of the inner disk is, however, not observationally constrained. Here, we simply assume that the inner disk mass is proportional to the total disk mass. The amount of the excess should depend on the disk inclination and the shape of the inner rim of the disk \citep{dullemond_inner_2010}. The connection between $f^{\rm ex}$ and the disk mass has not yet been understood well. For sake of simplicity, we assume that $f^{\rm ex}$ is given by an unary function of $m_{\rm d}$. The infrared excess $f^{\rm ex}(m_{\rm d})$ should be a non-decreasing function with $m_{\rm d}$. The threshold mass $m^{\rm th}_{\rm d}$ is defined by $f^{\rm ex}(m^{\rm th}_{\rm d}) = f^{\rm th}$. The disk is detected in the infrared when $m_{\rm d} \geq m^{\rm th}_{\rm d}$, while infrared excess is not detectable when $m_{\rm d} < m^{\rm th}_{\rm d}$. In this formalism, we do not specify the relationship between $f^{\rm ex}$ and $m_{\rm d}$. The presence or absence of the disk is determined only by $m_{\rm d}$, \textit{irrespective} of the functional form of $f^{\rm ex}(m_{\rm d})$. The infrared excess $f^{\rm ex}$ may depend on several parameters such as the disk inclination. Justification of this simple formulation is discussed in Section~\ref{sec:dis:justification}.

The disk fraction at the age of $t$ is calculated by
\begin{equation}
  \label{eq:diskfraction}
    {\mathcal F}(t)
    = \int_{m^{\rm th}_{\rm d}}^{\infty} \phi_t(m) {\rm d}m.
\end{equation}
Then, by substituting Equations~(\ref{eq:massfunction_logn}), (\ref{eq:disk_dissipation}), and (\ref{eq:diskmassfunction_with_age}) in Equation~(\ref{eq:diskfraction}), the disk fraction ${\mathcal F}(t)$ becomes
\begin{equation}
  \label{eq:diskfraction_reduced}
  \mathcal{F}\left(t; m^{\rm th}_{\rm d}, \mu, \sigma\right)
  = \frac{1}{2}
  {\rm erfc} \left[\frac{
    \log\mathcal{M}\left(-t,m^{\rm th}_{\rm d}\right)
    -\log\mu
  }{\sqrt{2}\sigma}\right],
\end{equation}
where ${\rm erfc}(x)$ is the complimentary error function. Since ${\rm erfc}(x)$ is a decreasing function on $x$, the disk fraction, ${\mathcal F}\left(t, m^{\rm th}_{\rm d}, \mu, \sigma \right)$, decreases as $t$ increases. The DTDF is defined as a typical timescale of the decrease in Equation~(\ref{eq:diskfraction_reduced}).

\section{Discussion}
\label{sec:discussion}
\begin{figure}
  \centering
  \includegraphics[width=\linewidth]{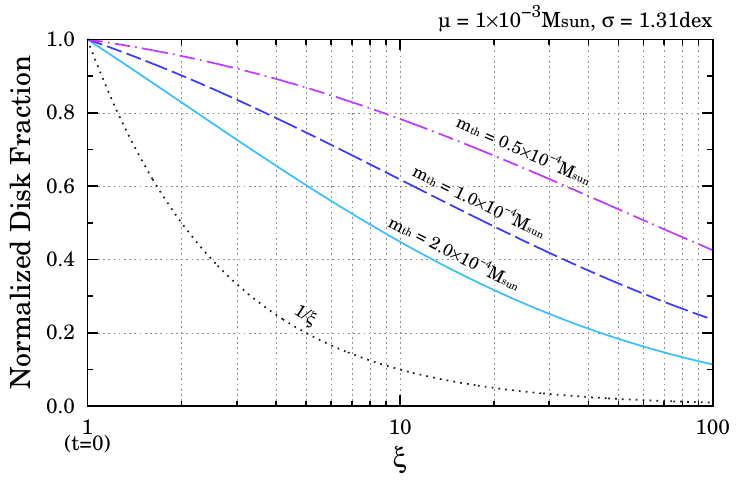}
  \caption{Time evolution of the disk fraction against ${\xi}$. Note that the horizontal axis is ${\xi}$, not the time $t$, because the variation of the disk fraction can be described in terms of ${\xi}$ in a more general manner than of $t$ (\textit{see} text). The disk fractions normalized at age zero are plotted for $m^{\rm th}_{\rm d} = 2.0{\times}10^{-4},\,1.0{\times}10^{-4},$ and $0.5{\times}10^{-4}\,M_{\odot}$ by the solid, dashed, and dot-dashed lines, respectively. The median and variance of the DMF are fixed at $10^{-3}\,M_{\odot}$ and $1.31\,$dex, respectively. The dotted curve shows the reciprocal of ${\xi}$, indicating the decrease in the individual disk mass (\textit{see}, text).}
  \label{fig:diskfraction_Tx}
\end{figure}

\subsection{Behavior of Disk Fraction}\label{sec:dis:behavior}
Since the complimentary error function ${\rm erfc}(x)$ is a monotonically decreasing function, the disk fraction will decrease with increasing $\mathcal{M}\left(-t,m_{\rm d}(0)\right)$, which increases monotonically with $t$. Equation (\ref{eq:diskfraction_reduced}) shows that the central loci of the DMF $\mu$ shifts the disk function along the axis of age. The width of the DMF $\sigma$ normalizes the argument of the complimentary error function in Equation~(\ref{eq:diskfraction_reduced}). As $\sigma$ increases by a factor of $x$, the time variation of the disk fraction is decelerated by a factor of $x$. This suggests that the broadness of the DMF has a large impact on the evolution of the disk fraction.

To investigate a typical decreasing timescale of Equation~(\ref{eq:mt_solved}), we define the time $T(\xi)$ such that the disk with the initial mass of ${\xi}m^{\rm th}_{\rm d}$ loses its mass down to $m^{\rm th}_{\rm d}$ at $t = T(\xi)$:
\begin{equation}
  \label{eq:definition_Tx}
  \mathcal{M}\left(T(\xi),{\xi}m^{\rm th}_{\rm d}\right) = m^{\rm th}_{\rm d}.
\end{equation}
{The duration means an observational lifetime of the disk with the initial mass of ${\xi}m^{\rm th}_{\rm d}$.} Equation~(\ref{eq:definition_Tx}) is equivalent to
\begin{align}
  \label{eq:time_Tx}
  -T(\xi) + \mathcal{Z}&\left({\xi}m^{\rm th}_{\rm d}\right)
  = \mathcal{Z}\left(m^{\rm th}_{\rm d}\right)
  \nonumber \\
  & ~~\iff~~
  T(\xi) = \mathcal{Z}\left({\xi}m^{\rm th}_{\rm d}\right)
  - \mathcal{Z}\left(m^{\rm th}_{\rm d}\right).
\end{align}
$T(\xi)$ increases with ${\xi}$ since $\mathcal{Z}(m)$ is an increasing function. When ${\xi}$ is unity, $T(\xi)$ is equal to zero. The time $T(\xi)$ is regarded as a typical timescale for the disk mass to decrease by a factor of ${\xi}$. By substituting Equation~(\ref{eq:time_Tx}), the following relationship is obtained:
\begin{equation}
  \label{eq:m0_Tx}
  \mathcal{M}\left(-T(\xi),m^{\rm th}_{\rm d}\right) =
  \mathcal{Z}^{-1}\left(T(\xi) + \mathcal{Z}(m^{\rm th}_{\rm d})\right)
  = {\xi}m^{\rm th}_{\rm d}.
\end{equation}
The disk fraction at $T(\xi)$ is obtained as
\begin{equation}
  \label{eq:diskfraction_reduced_Tx}
  \mathcal{F}\left(T(\xi); m^{\rm th}_{\rm d}, \mu, \sigma\right)
  = \frac{1}{2}
  {\rm erfc} \left[\frac{
      \log\left({\xi}m^{\rm th}_{\rm d}/\mu\right)
    }{\sqrt{2}\sigma}\right].
\end{equation}
The disk fraction decreases with ${\xi}$ as well as $T({\xi})$. Equation~(\ref{eq:diskfraction_reduced_Tx}) does not include $T(\xi)$ explicitly, indicating that the relationship between the disk fraction and ${\xi}$ does not depend on the functional form of $T(\xi)$. Thus, the evolution of the disk fraction can be described in a general manner in terms of ${\xi}$. Figure~\ref{fig:diskfraction_Tx} shows the evolution of the disk fractions normalized at age zero against ${\xi}$. The DMF of the Taurus cloud is assumed to be a representative one, so that $\mu$ and $\sigma$ are fixed at $1.0{\times}10^{-3}\,M_\odot$ and $1.31\,$dex, respectively \citep[the ``full'' sample in][]{andrews_circumstellar_2005}\footnote{Although \citet{andrews_circumstellar_2005} have reported that the median and the variance are $5{\times}10^{-3}\,M_{\odot}$ and $0.50\,$dex for the ``detection'' sub-sample, we adopt the values for the ``full'' sample because non-detected samples should be properly taken into account.}. By comparing sub-mm and near-infrared observations, \citet{andrews_circumstellar_2005} suggest that the near-infrared ($K_s{-}L$) color excess method can detect the protoplanetary disk with a mass down to $\sim 10^{-4}\,M_\odot$. The disk fractions are plotted for $m^{\rm th}_{\rm d} = 0.5{\times}10^{-4},\,1.0{\times}10^{-4},$ and $2.0{\times}10^{-4}\,M_{\odot}$. The gray dotted curve shows the reciprocal of ${\xi}$, representing the dissipation of the disk mass. If the disk fraction decreases with the same timescale as the disk mass, the curve should follow the dotted curve. When the disk fraction decreases slower than the disk mass, the curve will deviate upward, and vice versa. All the disk fractions deviate upward in Figure~\ref{fig:diskfraction_Tx}. An ${\rm e}$-folding time of the individual disk mass dissipation estimated by $T({\rm e})$, which corresponds to the MDTID. On the other hand, the disk fraction for $m^{\rm th}_{\rm d} = \,1.0{\times}10^{-4}$ does not decrease by a factor of ${\rm e}$ until $\xi$ increases to about $40$, suggesting that the DTDF is as long as $T(40)$. We found that the disk fraction decreases as fast as the disk mass only when $m^{\rm th}_{\rm d} \sim 2{\times}10^{-1}\,M_{\odot}$. Such an extreme case is excluded from observations. Figure~\ref{fig:diskfraction_Tx} indicates that the disk fraction generally decreases slower than the disk mass when a typical initial DMF and a detection threshold are assumed.

\subsection{Case Study: Exponential Decay}\label{sec:dis:exponential}

\begin{figure}
  \centering
  \includegraphics[width=\linewidth]{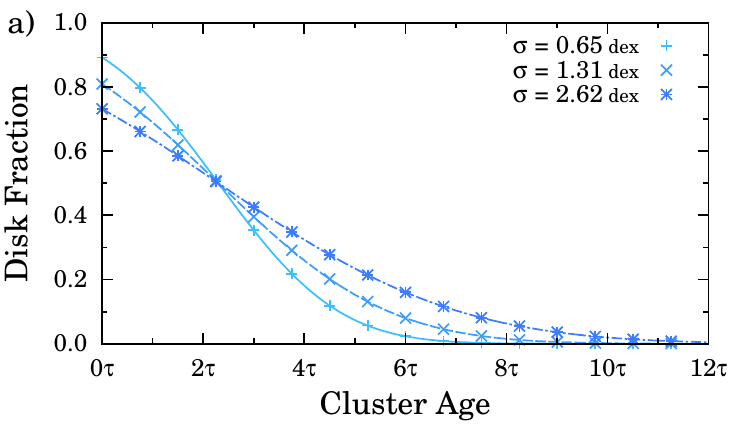}
  \includegraphics[width=\linewidth]{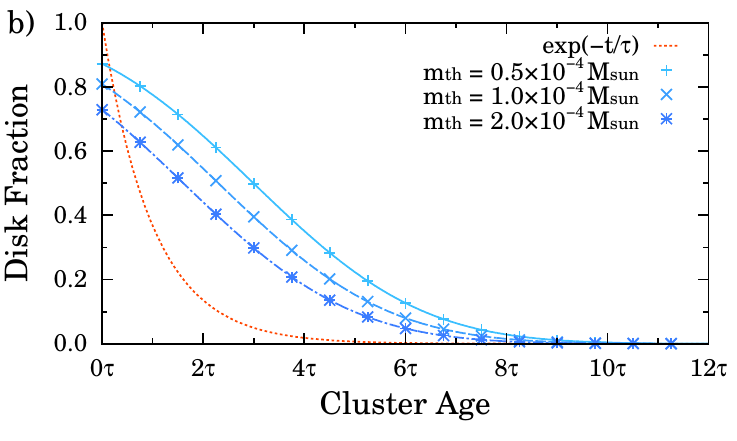}
  \caption{\textbf{a)} Disk fractions for different $\sigma$'s. \textbf{b)} Disk fractions for different $m^{\rm th}_{\rm d}$'s. The horizontal axis is the cluster age $t$ measured by $\tau$. The red dot-dashed line shows n exponential decay with a time constant of $\tau$.}
  \label{fig:diskfraction}
\end{figure}

To quantitatively compare the MDTID and DTDF, we consider a special case of $\zeta(m) = \tau/m$, which corresponds to an exponential decay. Equation~(\ref{eq:mt_solved}) is reduced to
\begin{equation}
  \label{eq:disk_dissipation_exp}
  m_{\rm d}(t) = m_{\rm d}(0) {\rm e}^{-\frac{t}{\tau}}.
\end{equation}
Here $\tau$ is equivalent to $T({\rm e})$. In this section, we define the MDTID by the ${\rm e}$-folding time $\tau$. With the assumption of $\zeta(m) = \tau/m$, the DMF at $t$ is given by
\begin{align}
  \label{eq:massfunction_exp}
  \phi_t&\left( m_{\rm d} \right) {\rm d}m_{\rm d}
  = \nonumber\\
  & \frac{1}{m_{\rm d}\sqrt{2\pi}\sigma}
  \exp\left[
    -\frac{1}{2}
    \left(
      \frac{
      \log{m_{\rm d}}{-}\log{\mu {\rm e}^{-\frac{t}{\tau}}}
      }{\sigma}
    \right)^2
  \right]{\rm d}m_{\rm d}.
\end{align}
Here, the DMF remains a log-normal distribution. The median of the DMF at $t$ is given by $\mu{\rm e}^{-\frac{t}{\tau}}$, indicating that the location of the DMF shifts with the time constant of $\tau$, while the scale parameter of the log-normal distribution, $\sigma$, is constant in time. By applying Equation~(\ref{eq:disk_dissipation_exp}), the disk fraction is given by
\begin{equation}
  \label{eq:diskfraction_reduced_exp}
  {\mathcal F}\left(t, m^{\rm th}_{\rm d}, \mu, \sigma \right)
  = \frac{1}{2}
  {\rm erfc}\left[
    \frac{t/\tau + \log\left({m^{\rm th}_{\rm d}}/\mu\right)}{\sqrt{2}\sigma}
  \right].
\end{equation}
Figure \ref{fig:diskfraction} shows the time variation of the disk fractions for different DMFs. The horizontal axis denotes the cluster age $t$ measured by $\tau$. The median mass of DMF is set to be $1.0{\times}10^{-3}\,M_\odot$ \citep{andrews_circumstellar_2005}. Figure~\ref{fig:diskfraction}a shows the results for different DMFs, $\sigma = 0.65$, $1.31$, and $2.62\,$dex, where $m^{\rm th}_{\rm d}$ is fixed at $10^{-4}\,M_\odot$. The figure indicates that the disk fraction decreases slower as the disk mass function becomes broader, because $t$ is normalized by $\sigma\tau$ in Equation~(\ref{eq:diskfraction_reduced_exp}). The disk fraction is not unity even at age zero, simply because some disks do not have a large amount of initial mass enough to be detected. Figure~\ref{fig:diskfraction}b shows the disk fractions for different detection thresholds, $m^{\rm th}_{\rm d} = 0.5{\times}10^{-4}$, $1.0{\times}10^{-4}$, and $2.0{\times}10^{-4}\,M_\odot$. The disk fraction shifts to the left with increasing $m^{\rm th}_{\rm d}$. The parameter $\left(m^{\rm th}_{\rm d}/\mu\right)$ affects the disk fraction at age zero, but does not change the slope of the disk fraction.

The red dot-dashed line in Figure~\ref{fig:diskfraction}a shows an exponential decay curve with the time constant of $\tau$: $\exp(-t/\tau)$. All the disk fractions in Figure~\ref{fig:diskfraction}a decrease more slowly than the exponential curve. The ${\rm e}$-folding time of the disk fractions ranges between $2\tau$ and $4\tau$, depending on $m^{\rm th}_{\rm d}$. This is consistent with the discussion in Section~\ref{sec:dis:behavior}: the disk fraction decreases generally more slowly than the disk dissipates. Figure~\ref{fig:diskfraction} implies that the difference between the MDTID and DTDF is not negligible.

\begin{figure}
  \centering
  \includegraphics[width=\linewidth]{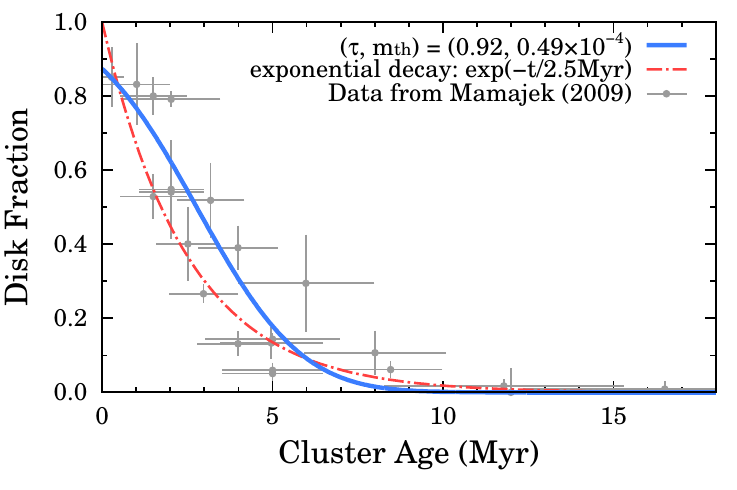}
  \caption{Comparison with observed disk fractions. The gray filled circles with errors are observed disk fractions from \citet{mamajek_initial_2009}. The blue solid line shows the disk fraction with the best-fit parameters. The red dot-dashed line shows an exponential decay with a time constant of $2.5\,$Myr.}
  \label{fig:observation}
\end{figure}

As a demonstration, we calculate the difference between DTDF and MDTID using a real data set. Since the present model is highly simplified, this practice is to demonstrate the degree of effect of the initial DMF on the estimate of the MDTID. We do not intend to derive an accurate estimate of the MDTID. Figure~\ref{fig:observation} shows that the observed disk fractions of a number of star clusters collected in \citet{mamajek_initial_2009} and references therein. In the current model, the disk fraction is given by Equation~(\ref{eq:diskfraction_reduced_exp}). The location and width of the initial DMF, $\mu$ and $\sigma$, are fixed at $1.0{\times}10^{-3}\,M_\odot$ and $1.31\,$dex, respectively \citep{andrews_circumstellar_2005}. The other two parameters, $\tau$ and $m^{\rm th}_{\rm d}$, are estimated by a maximum likelihood method. The likelihood function is defined as
\begin{equation}
  \label{eq:likelihood}
  \mathcal{L}(\tau,\gamma) \propto
  \prod_i \exp\left[
    -\frac{1}{2}
    \left(
      \frac{f_i - \mathcal{F}\left(t_i;\tau,m^{\rm th}_{\rm d}\right)}
      {\Delta f_i}
    \right)^2
  \right],
\end{equation}
where $f_i$ and $\Delta f_i$ are the observed disk fraction and its uncertainty, and $t_i$ is the cluster age. The parameters, $\tau$ and $m^{\rm th}_{\rm d}$, are estimated by maximizing Equation~(\ref{eq:likelihood}). The errors in $\tau$ and $m^{\rm th}_{\rm d}$ are estimated by parametric bootstrapping. The best fit parameters are $\tau = 0.92^{{+}0.39}_{{-}0.28}\,$Myr and $m^{\rm th}_{\rm d} = 0.48^{{+}0.71}_{{-}0.26}{\times}10^{-4}\,M_\odot$. Even though the model is simple and ignores many effects, the detection threshold presented above is consistent with the value reported by \citet{andrews_circumstellar_2005} (${\sim}10^{-4}\,M_\odot$). The blue line with the crosses in Figure \ref{fig:observation} shows the fitted curve. \citet{mamajek_initial_2009} suggests that the evolution of the observed disk fraction is well represented by an exponential decay with the time constant of $2.5\,$Myr, which is shown by the red dot-dashed line. The decrease in the disk fraction is well approximated by $\exp(-t/T)$, where $T = 2.5\,$Myr, suggesting that DTDF is about $2.5\,$Myr. The present result, however, suggests that MDTID $\tau$ is as short as $0.9\,$Myr.

\subsection{Justification of Model}\label{sec:dis:justification}
The discussion in Section~\ref{sec:dis:behavior} is applied to a general case, where various physical processes --- such as binarity, variable X-ray flux, and photo-evaporation --- can contribute to the dissipation. It assumes that the initial DMF is given by a log-normal function, but does not have a constraint on the DMF at a given time. Although the present model is rather robust in this sense, it is simplified and based on several assumptions. Effects of the simplification on the results are discussed.

In the present model, the stars in the cluster are assumed to be coeval. The star formation is, however, not instantaneous. \citet{getman_age_2014} investigate the age of individual YSOs in massive star forming regions and show that the spread of the stellar ages in a cluster is about $1$--$2\,$Myr, comparable with the DTDF ($\sim 2.5\,$Myr). The age spread within a cluster should not be neglected in the discussion of the disk fraction. At a given epoch, YSOs formed earlier have lost more mass than those formed later, and vice versa. {Thus, the apparent width of the DMF is broadened if the age distribution is taken into account in the present model. While the width of the DMFs has been observationally measured \citep[e.g.,][]{andrews_circumstellar_2005}, the intrinsic width of the DMF and the broadening effect due to the age spread are not distinguished. The observationally estimated width of the DMF should include the effect of the age distribution. Since we employ the observationally estimated width of the DMF \citep{andrews_circumstellar_2005} in our discussion, the present results indirectly take account of the effect of the age distribution.}

The present model assumes that the mass of the inner disk is proportional to the total disk mass. This is not observationally confirmed, because the inner disk is optically thick in the infrared and the infrared excess does not reflect directly the amount of the inner disk mass. The present results, however, mainly depend on the width of the initial DMFs. The results are expected to be valid as long as the distribution of the inner disk mass is relatively as wide as the DMFs observed in the sub-mm wavelength.

The present model does not include any inclination effect on the infrared excess, although the infrared excess $f^{\rm ex}$ should be changed with inclination. The relationship between $f^{\rm ex}$ and the inclination angle may be complicated. Since the near-infrared emission is mainly from the inner rim of the dusty disk \citep{dullemond_passive_2001,natta_reconsideration_2001,muzerolle_unveiling_2003}, the effect of the inclination angle can heavily depend on the shape of the inner rim \citep{dullemond_inner_2010}. If the inner rim has a round shape \citep{isella_shape_2005}, the effect of the inclination on $f^{\rm ex}$ can be small. Observationally the infrared excess does not show any strong correlation between the inclination \citep{dullemond_inner_2010}, suggesting that the inclination effect on the present result is not significant. Although the inclination significantly changes the infrared excess $f^{\rm ex}$, it is hard to distinguish the contribution from the inclination and the disk mass unless the inclination angle is determined. The effect of the inclination can be practically taken in the present model by broadening the initial DMF.

The relationship between the infrared excess $f^{\rm}$ and the disk mass, which is important to connect the initial DMF with the disk fraction, remains to be understood. We assume that the infrared excess $f^{\rm}$ does not decrease as the disk mass increases. In the present model, the infrared excess $f^{\rm ex}$ is given by an unary function of the disk mass. The amount of the infrared excess is closely related to the shape of the inner rim \citep{dullemond_inner_2010,isella_shape_2005}. The shape of the rim depends on the pressure scale height at the inner rim $H_{p,{\rm rim}}$, which is proportional to $1/\sqrt{M_\star}$, where $M_{\star}$ is the stellar mass \citep{isella_shape_2005}. Therefore, the neglect of the mass dependence can make a moderate effect on the result and should be taken into account in the next step. The effect on the present results due to this simplification is not expected to be severe as long as the mass range is sufficiently narrow.

In Equation~(\ref{eq:disk_dissipation}), we assume that the decrease in the disk mass is given by an unary function of $m_{\rm d}$. This assumption can be rephrased as disks with the same disk mass have the same lifetime. The decreasing rate of the disk mass should depend on the radiation from the central star and neighborhood stars. The assumption can become invalid if the mass range is not sufficiently narrow or the radiation field significantly changes within the cluster. Consequently, the lifetime of the disk can be extended or shortened. The effects of the mass range and the radiation field of neighborhood stars can be qualitatively estimated by broadening the width of the initial DMF. As shown in Equation~(\ref{eq:diskfraction_reduced}), the disk fraction decreases more slowly than the disk mass as the initial DMF becomes wider. Although the present model does not correctly take account of the effects of the mass range and the radiation field from neighborhood stars, those effects should not change the conclusion of the present result qualitatively.

\subsection{Implications of Results}\label{sec:dis:implications}
As shown in Figure \ref{fig:diskfraction}, the apparent evolution of the disk fraction depends on the shape of the initial DMF. The disk fractions of clusters with different initial DMFs should follow different curves. In most observational studies, the disk fractions are assumed to follow a single curve \citep[e.g.,][]{haisch_disk_2001,yasui_lifetime_2009,yasui_rapid_2014,ribas_disk_2014}. In other words, they assume that the initial DMFs are assumed to be the same. However, the DMF has so far been obtained for only a few clusters, e.g., in the Taurus, Ophiuchus, and Orion regions \citep[e.g.,][]{williams_protoplanetary_2011}. Those observations suggest that the DMF can be approximated fairly well by a single log-normal function, while \citet{mann_protoplanetary_2015} suggest that the DMF of NGC\,2024 is possibly top-heavy in comparison with that of the Taurus cloud and Orion Nebula Clusters. If the DMF is top-heavy in the sample clusters, the DTDF will become even longer than the present estimate. Only by investigating the evolution of the disk fraction, it is not possible to disentangle the variation in the MDTID and the variation in the initial DMFs. Further investigation about the variation in the initial DMF is needed.

{Previous studies have been compared the disk fractions of different clusters and discussed the DTDF \citep[e.g.,][]{haisch_disk_2001,yasui_short_2010}. The present results suggest that the apparent evolution of the disk fraction depends on the threshold mass. The disk detected in the near-infrared is expected to be fully optically thick. The amount of excess $f^{\rm ex}$ in the near-infrared is insensitive to the disk mass in the optically thick phase \citep{wood_infrared_2002}. The choice of the detection threshold $(f^{\rm th})$ will not matter. In general, the transition from a thick to thin disk occurs in a small mass range at the near-infrared wavelength. The fraction of disks in the transition in a cluster is small \citep[e.g.,][]{skrutskie_sensitive_1990,wolk_search_1996,cieza_spitzer_2007}. Thus, the dependence of the threshold mass $m^{\rm th}_{\rm d}$ is negligible. At wavelengths longer than $25\,\mu$m, the disk can be detected in the optically thin phase and the excess becomes sensitive to the disk mass \citep{wood_infrared_2002}. The difference in the detection threshold may have to be taken into account.}

\section{Conclusion}
\label{sec:conclusion}
The disk fraction, which is the fraction of stars with disks in a young cluster, is widely used to observationally investigate the lifetime of the protoplanetary disk. The time evolution of the disk fraction should depend on the disk mass function (DMF) at age zero. We discuss a simple model to analytically investigate the relationship between the mass dissipation timescale of an individual disk (MDTID) and the decay timescale of the disk fraction (DTDF).

In the present model, the evolution of the disk fraction can be described by the detection threshold of the disk mass ($m^{\rm th}_{\rm d}$), and the loci ($\mu$) and the dispersion ($\sigma$) of the initial DMF. The DMF of the Taurus cluster has $\mu \simeq 10^{-3}\,M_{\odot}$ and $\sigma \simeq 1.31\,{\rm dex}$ \citep{andrews_circumstellar_2005}. Using the near-infrared color excess method, the disk with the mass of ${\sim}10^{-4}\,M_{\odot}$ can be detected \citep{andrews_circumstellar_2005}. Assuming these parameters, the present model indicates that the disk fraction generally decreases more slowly than the disk mass, suggesting that the DTDF is longer than the MDTID. Given that the disk mass dissipates exponentially, the difference between the DTDF and the MDTID is suggested not to be negligible. Comparison with the observational data \citep{mamajek_initial_2009} suggests that the MDTID is as small as $1\,$Myr even if the disk fraction exponentially decreases with the time constant of $2.5\,$Myr, corresponding to the DTDF. Although the present model is simple and primitive, we are confident that the present results are qualitatively valid.

The present results suggest that the evolution of the disk fraction depends on the shape of the initial DMF. The variation in the DMF remains to be understood. Further observational investigation is needed. {Although the apparent evolution of the disk fraction may depend on the detection threshold mass, the matching of the detection threshold does not matter unless the disk fraction is measured at a longer (${\gtrsim}25\,\mu$m) wavelength.}

The study of the protoplanetary disk is being accelerated by the advent of \textit{ALMA}, which can resolve the disk structure and precisely measure the mass of the disk. However, until the disk mass is accurately measured for a sufficiently large sample of disks in various environments, statistical approach is still important for the study of the disk evolution. The disk fraction remains to be efficient and useful if the relationship between the DTDF and the MDTID is properly appreciated.

\begin{ack}
This research is supported in part by Grants-in-Aid for Scientific Research (25-8492, 23103004, 26247074, and 26800094) by the Japan Society of Promotion of Science.

\end{ack}

\bibliographystyle{apj}

\begin{thebibliography}{37}
\expandafter\ifx\csname natexlab
\endcsname\relax\def\natexlab#1{#1}\fi
\bibitem[Andrews \& Williams(2005)]{andrews_circumstellar_2005}Andrews, S.~M., \& Williams, J.~P. 2005, The Astrophysical Journal, 631, 1134
\bibitem[Andrews \& Williams(2007)]{andrews_submillimeter_2007}---. 2007, The Astrophysical Journal, 671, 1800
\bibitem[Armitage(2011)]{armitage_dynamics_2011}Armitage, P.~J. 2011, Annual Review of Astronomy and Astrophysics, 49, 195
\bibitem[Armitage {et~al.}(2003)]{armitage_dispersion_2003}Armitage, P.~J., Clarke, C.~J., \& Palla, F. 2003, Monthly Notices of the Royal  Astronomical Society, 342, 1139
\bibitem[Bachiller(1996)]{bachiller_bipolar_1996}Bachiller, R. 1996, Annual Review of Astronomy and Astrophysics, 34, 111
\bibitem[Carpenter {et~al.}(2006)]{carpenter_evidence_2006}Carpenter, J.~M., Mamajek, E.~E., Hillenbrand, L.~A., \& Meyer, M.~R. 2006, The  Astrophysical Journal Letters, 651, L49
\bibitem[Cieza {et~al.}(2007)]{cieza_spitzer_2007}Cieza, L., {et~al.} 2007, The Astrophysical Journal, 667, 308
\bibitem[Dullemond {et~al.}(2001)]{dullemond_passive_2001}Dullemond, C.~P., Dominik, C., \& Natta, A. 2001, The Astrophysical Journal,  560, 957
\bibitem[Dullemond \& Monnier(2010)]{dullemond_inner_2010}Dullemond, C.~P., \& Monnier, J.~D. 2010, Annual Review of Astronomy and  Astrophysics, 48, 205
\bibitem[Getman {et~al.}(2014)]{getman_age_2014}Getman, K.~V., {et~al.} 2014, The Astrophysical Journal, 787, 108
\bibitem[Haisch {et~al.}(2001)]{haisch_disk_2001}Haisch, K.~E., Lada, E.~A., \& Lada, C.~J. 2001, Astrophysical Journal, 553,  L153
\bibitem[Hern{\'a}ndez {et~al.}(2005)]{hernandez_herbig_2005}Hern{\'a}ndez, J., {et~al.} 2005, The Astronomical Journal, 129, 856
\bibitem[Hern{\'a}ndez {et~al.}(2008)]{hernandez_spitzer_2008}---. 2008, The Astrophysical Journal, 686, 1195
\bibitem[Ida \& Lin(2004)]{ida_toward_2004}Ida, S., \& Lin, D. N.~C. 2004, The Astrophysical Journal, 604, 388
\bibitem[Isella \& Natta(2005)]{isella_shape_2005}Isella, A., \& Natta, A. 2005, Astronomy and Astrophysics, 438, 899
\bibitem[Kennedy \& Kenyon(2009)]{kennedy_stellar_2009}Kennedy, G.~M., \& Kenyon, S.~J. 2009, The Astrophysical Journal, 695, 1210
\bibitem[Kenyon \& Hartmann(1995)]{kenyon_pre-main-sequence_1995}Kenyon, S.~J., \& Hartmann, L. 1995, The Astrophysical Journal Supplement  Series, 101, 117
\bibitem[Luhman {et~al.}(2010)]{luhman_disk_2010}Luhman, K.~L., {et~al.} 2010, The Astrophysical Journal Supplement Series, 186,  111
\bibitem[Mamajek(2009)]{mamajek_initial_2009}Mamajek, E.~E. 2009, in {EXOPLANETS} {AND} {DISKS}: {THEIR} {FORMATION} {AND}  {DIVERSITY}, Vol. 1158 (Melville, NY, USA: American Institute of Physics),  3--10
\bibitem[Mann {et~al.}(2015)]{mann_protoplanetary_2015}Mann, R.~K., {et~al.} 2015, ApJ, in press
\bibitem[Mann \& Williams(2010)]{mann_submillimeter_2010}Mann, R.~K., \& Williams, J.~P. 2010, The Astrophysical Journal, 725, 430
\bibitem[Muzerolle {et~al.}(2003)]{muzerolle_unveiling_2003}Muzerolle, J., Calvet, N., Hartmann, L., \& D'Alessio, P. 2003, The  Astrophysical Journal Letters, 597, L149
\bibitem[Natta {et~al.}(2001)]{natta_reconsideration_2001}Natta, A., {et~al.} 2001, Astronomy and Astrophysics, 371, 186
\bibitem[Owen {et~al.}(2011)]{owen_protoplanetary_2011}Owen, J.~E., Ercolano, B., \& Clarke, C.~J. 2011, Monthly Notices of the Royal  Astronomical Society, 412, 13
\bibitem[Ribas {et~al.}(2014)]{ribas_disk_2014}Ribas, {\'A}., Mer{\'i}n, B., Bouy, H., \& Maud, L.~T. 2014, Astronomy and  Astrophysics, 561, 54
\bibitem[Skrutskie {et~al.}(1990)]{skrutskie_sensitive_1990}Skrutskie, M.~F., {et~al.} 1990, The Astronomical Journal, 99, 1187
\bibitem[Strom {et~al.}(1989)]{strom_circumstellar_1989}Strom, K.~M., {et~al.} 1989, The Astronomical Journal, 97, 1451
\bibitem[Takagi {et~al.}(2010)]{takagi_age-determination_2010}Takagi, Y., Itoh, Y., \& Oasa, Y. 2010, Publications of the Astronomical  Society of Japan, 62, 501
\bibitem[Takagi {et~al.}(2011)]{takagi_age_2011}Takagi, Y., Itoh, Y., Oasa, Y., \& Sugitani, K. 2011, Publications of the  Astronomical Society of Japan, 63, 677
\bibitem[Williams \& Cieza(2011)]{williams_protoplanetary_2011}Williams, J.~P., \& Cieza, L.~A. 2011, Annual Review of Astronomy and  Astrophysics, 49, 67
\bibitem[Wolk \& Walter(1996)]{wolk_search_1996}Wolk, S.~J., \& Walter, F.~M. 1996, The Astronomical Journal, 111, 2066
\bibitem[Wood {et~al.}(2002)]{wood_infrared_2002}Wood, K., {et~al.} 2002, The Astrophysical Journal, 567, 1183
\bibitem[Wyatt(2008)]{wyatt_evolution_2008}Wyatt, M.~C. 2008, Annual Review of Astronomy and Astrophysics, 46, 339
\bibitem[Wyatt {et~al.}(2007)]{wyatt_origin_2007}Wyatt, M.~C., Clarke, C.~J., \& Greaves, J.~S. 2007, Monthly Notices of the  Royal Astronomical Society, 380, 1737
\bibitem[Yasui {et~al.}(2014)]{yasui_rapid_2014}Yasui, C., Kobayashi, N., Tokunaga, A.~T., \& Saito, M. 2014, Monthly Notices  of the Royal Astronomical Society, 442, 2543
\bibitem[Yasui {et~al.}(2009)]{yasui_lifetime_2009}Yasui, C., {et~al.} 2009, The Astrophysical Journal, 705, 54
\bibitem[Yasui {et~al.}(2010)]{yasui_short_2010}---. 2010, The Astrophysical Journal Letters, 723, L113
\end{thebibliography}

\end{document}